\documentclass[a4paper,aps,prl,reprint,superscriptaddress,twocolumn,floatfix]{revtex4-1}

\pdfoutput=1

\usepackage{amsmath}
\usepackage{graphicx}
\usepackage{color}
\usepackage{amssymb}
\usepackage{hyperref}
\usepackage[caption=false]{subfig}

\begin{document}

\newcommand{\beq}{\begin{equation}}
\newcommand{\eeq}{\end{equation}}
\newcommand{\beqa}{\begin{eqnarray}}
\newcommand{\eeqa}{\end{eqnarray}}
\newcommand{\ben}{\begin{enumerate}}
\newcommand{\een}{\end{enumerate}}
\newcommand{\hs}{\hspace{0.5cm}}
\newcommand{\vs}{\vspace{0.5cm}}

\title{Geometrical mutual information at the tricritical point \\ of the two--dimensional Blume--Capel model}

\author{Ipsita Mandal}
\affiliation{Perimeter Institute for Theoretical Physics, Waterloo, Ontario N2L 2Y5, Canada}

\author{Stephen Inglis}
\affiliation{Department of Physics and Arnold Sommerfeld
Center for Theoretical Physics, Ludwig-Maximilians-Universit\"at
M\"unchen, D-80333 M\"unchen, Germany}

\author{Roger G. Melko}
\affiliation{Perimeter Institute for Theoretical Physics, Waterloo, Ontario N2L 2Y5, Canada}
\affiliation{Department of Physics and Astronomy, University of Waterloo, Ontario, N2L 3G1, Canada}

\date{\today}

\begin{abstract}
The spin-1 classical Blume-Capel model on a square lattice 
is known to exhibit a finite-temperature phase transition
described by the tricritical Ising CFT in 1+1 space-time dimensions. 
This phase transition can be accessed with classical Monte Carlo simulations, 
which, via a replica-trick calculation, can be used to study the shape-dependence of the classical R\'enyi entropies
for a torus divided into two cylinders.
From the second R\'enyi entropy, we calculate the Geometrical Mutual Information (GMI) introduced by St\'ephan {\it et. al.}~
[Phys. Rev. Lett. 112, 127204 (2014)]
and use it to extract a numerical estimate for the value of the central charge near the tricritical point. 
By comparing to the known CFT result, $c=7/10$, we
demonstrate how this type of GMI calculation can be used to estimate the position of the tricritical point in the phase diagram.
\end{abstract}

\maketitle

{\em Introduction --}
It is now well-established that there is a deep connection between certain measurable thermodynamic quantities
and principles of information theory.  
Most straightforwardly, information can be quantified in terms of entropy, which can be defined from thermodynamic observables \cite{shannon,cardy}. 
For finite-temperature phase transitions, critical points are characterized by infinite
correlation lengths, indicative of the existence of long-range
channels for information transfer.  
It is interesting to ask whether observables derived from information quantities can be used to characterize
classical phase transitions.
Despite the answer being non-trivial, the R\'enyi entropies have recently been used to 
detect and classify phase transitions in a number of
classical systems \cite{stephen2013,stephan2014,troyer,vidal,Alba1,Alba2,stephen2016,Johannes}.
In particular, a mutual information derived from the second R\'enyi entropy \cite{melko2010,Singh,WL} has been very successful in 
detecting finite-temperature critical points, even identifying their universality class
without any {\it a priori} knowledge of an order parameter or associated broken symmetry.

The utility of the second R\'enyi entropy was demonstrated in a striking way by the introduction of the 
``Geometrical Mutual Information" (GMI) by St\'ephan {\it et. al.} \cite{stephan2014}, where it was shown that a 
simple-to-implement classical Monte Carlo simulation of an Ising model at its phase transition
was capable of calculating the central charge $c$ of the associated 1+1-dimensional conformal field theory (CFT)
\cite{belavin,friedan,wilczek,kitaev,cardy}.
Most straightforwardly, if the simulation is tuned to twice the critical temperature $T_c$ (for the second R\'enyi entropy), then a 
simple finite-size scaling analysis is sufficient to extract $c$ using the functional form for the GMI obtained in
Ref.~\cite{stephan2014} for a general CFT.  
It follows that one may employ the GMI in the converse manner: knowing the expected theoretical value of the central charge $c$,
the GMI can be used to estimate the parameters which lead to criticality in a model.  
This may be useful when two or more parameters must be tuned to realize a phase transition, 
such as occurs in the case of the tricritical Ising transition, which is described by a minimal CFT with $c=0.7$ \cite{belavin,friedan}.

The simplest classical model that realizes a tricritical Ising point is the spin-1 Blume-Capel model.
The Hamiltonian of the model
 on a two-dimensional square lattice \cite{blume,capel} is given by
%%%%%%%%%%%%%
\beq 
H = -J \sum_{\langle i j \rangle} S^z_i S^z_j
+ D \sum_{  i  } ( S ^z_i) ^2   ,
\label{bc-model}
\eeq
%%%%%%%%%%%%%%%
where $S^z_i = \left( \pm 1, 0 \right) $ and $ \langle i j \rangle $ denotes nearest-neighbor sites.
Below, we have set the energy scale $J=1$.
The model has been shown to exhibit a tricritical point
described by tricritical Ising CFT \cite{balbao}, though it cannot be solved exactly (non-integrable) away from criticality. The position of the
tricritical point is non-universal and can only be determined numerically.
There have been
extensive studies in the literature using various sophisticated numerical techniques to pin down the values of the parameter $D$ and temperature $ T$ of the tricritical point \cite{burk,berker,burk2,ng,landau,selke,chak,beale,landau2,tucker,du,du2,plascak,silva,yusuf}. 
In this paper, we confirm through numerical calculation of the second R\'enyi entropy GMI that the central charge 
near the tricritical point is consistent with the value predicted by CFT, $c=7/10$.  In addition, we
demonstrate how {\it a priori} knowledge of this universal constant can be used in conjunction with the GMI to provide an estimate for the position of the
phase transition, without any reliance on the order parameter of the system.

%%%%%%%%%%%%%%%%%%%%%%%%%%%
{\em Method --}
%%%%%%%%%%%%%%%
Let us consider a classical spin system defined on a square lattice with Hamiltonian Eq.~\eqref{bc-model}.
We can partition the lattice into two regions, $A$ and $B$,
with the spin configurations within each subsystem labeled as
$i_A$ and $i_B$ respectively. The probability of state $i_A$ occurring in subregion $A$
is $p_{ i_ A }= \sum \limits_{i_B} p_{i_A ,i_B}$ , where $  p_{i_A ,i_B}
= e^{ - \beta E(i_A ,i_B )  } / Z [T]  $
is the probability of existence of any arbitrary state of the entire system, obtained from the Boltzmann distribution.
Here $E(i_A , i_B )$ is the energy
associated with the states $i_A$ and $i_B$, $\beta = {1} /{T}$,  and $Z[T] 
=  \sum \limits_{i_A ,i_B} e^{ - \beta E(i_A ,i_B )} $
is the partition function of $ A \cup B$. Now the second R\'enyi entropy for subregion $A$
is defined by \cite{melko2010}:
\begin{eqnarray}
\label{s2}
S_2 (A)&=& 
- \ln   \left ( \sum_{ i_A}  p_{ i_ A }^2   \right )  \nonumber \\
%%%%%%%%%%
&=& -\ln \left ( \sum_{ i_A}
\frac{  \sum \limits_{ i_B}  e^{ - \beta E(i_A ,i_B )}
 \,  \sum \limits_{ j_B}  e^{ - \beta E(i_A ,j_B )} ) }
{ Z^2 [T]}   \right ) \nonumber \\
%%%%%%%%%%
&=& -\ln \left ( Z[A,2,T] \right ) + 2 \ln \left ( Z[T] \right ) \,,
\end{eqnarray}
%%%%%%%%%%%%%%
where $Z [A, 2, T ] =  \sum \limits_{i_A,  i_B, j_B}  e^{ - \beta \lbrace E(i_A ,i_B ) + E(i_A ,j_B ) \rbrace}$
is the partition function of a new ``replicated'' system, such that the spins in subregion $A$ are constrained to be the same
in both the replicas, while the spins in subregion $B$ are unrestricted for the two copies. The first condition leads the spins in the bulk of subregion $A$ to behave as if their effective temperature is $T/2$ for local interactions. The R{\'e}nyi mutual information (RMI) can now be defined as the symmetric quantity:
\begin{eqnarray}
\label{rmi}
I_2 (A, B) &=& S_2 (A) + S_2 (B) - S_2 (A \cup B ) \nonumber \\
%%%%%%%%%%
&=& -\ln \left (
\frac{  Z[ A,2, T] \,  Z[B, 2, T]
}
{  Z^2 [T] \,  Z[T/2] }
\right) \,.
\end{eqnarray}
%%%%%%%%%%%%%
This quantity has been demonstrated useful in the past for detecting finite-temperature phase transitions
with great accuracy \cite{Singh,stephen2013,WL}.

In two dimensions, the RMI can be used to define a universal quantity
($ \mathcal {G}_2 $) called the geometric mutual information (GMI) \cite{stephan2014},
\begin{equation}
I_2 (A, B) = a_2 \,  L +  \mathcal {G}_2  + \mathcal{O} (1), \label{RMI}
\end{equation}
which is a function of the various aspect ratios in the system.
%%%%%%%%%%
Due to the symmetry of the RMI, all the bulk (``volume'') contributions occurring in the R{\'e}nyi entropy cancel. 
This leaves the ``area-law'' as the leading order term in Eq.~\eqref{RMI}, proportional to $L$, 
which is the length of the boundary between the subregions $A$ and $B$.
In two dimensions, the exact expression for
the partition function of a critical system can be found using CFT \cite{stephan2014,kleban1,kleban2,bondesan,cardy2,fradkin,zalatel,stephan2,eduardo,Singh,cardy3,affleck}. 
For free external boundary conditions at $T = 2 \, T_c $,
cutting an $L_x \times L_y$ system into two rectangular subregions $L_A \times L_y$
and $ L_B \times L_y$, the expression for GMI is given by \cite{stephan2014}:
%%%%
\begin{equation}
\label{gmi}
\mathcal  G_2 = \frac{c} {2}
\ln \left(
\frac { f (L_A / L_x)  \, f (L_B / L_x) }
{ \sqrt{ L_x }  \, f (L_y / L_x) }
\right ) .
\end{equation}
Here, $c$ is the central charge of the associated CFT description of the critical point appearing at temperature $T_c$, and
$f (u ) = \eta (i \, u)$ (where $\eta $ is the standard Dedekind eta function \cite{stegun}). 
This allows us to define a finite-size scaling procedure to extract $c$ from numerical calculations 
of the GMI.

We compute the GMI using Monte Carlo simulations and the transfer-matrix ratio trick
for classical systems \cite{gelman1998,tommaso,graph-theory}, using the formula
%%%%
\begin{eqnarray}
\label{ratio}
&& \frac{  Z [A,2, T] }   {  Z^2 [T] }
= \prod \limits_{i=0}^{N-1}  \frac{  Z [A_{ i+1 } ,2, T] }   {   Z [A_i ,2, T] } \,,\nonumber \\
%%%%%%%%%%
&& Z [A_0 ,2, T]   = Z^2 [T] \,.
\end{eqnarray}
%%%%%%%%%%%%%%%%
Here, $A_i$ 
denotes a series of $N$ blocks of increasing size, the consecutive blocks differing by a
one-dimensional strip of spins running parallel to the boundary separating $A$ and $B$, with $ A_0 $ being the empty region and $A_N = A$.
The algorithm is well documented in Ref.~\cite{stephan2014}.  In addition to the procedure presented there, we combine parallel tempering to ensure that the states used to estimate the ratios of the partition functions, $
\Big \lbrace \frac{  Z [A_{ i+1 } ,2, T] }   {   Z [A_i ,2, T] }  \Big \rbrace $ , are efficiently sampled.
This is important when trying to accurately locate the tricritical point in the Blume-Capel model where critical slowing down can bias results from Monte Carlo simulations.
In addition, having results over a grid of model parameters allows us to examine the quality of the fit to the universal shape function to compare to previous estimates for the tricritical temperature $T_{tc}$ and the coupling constant $D_{ tc } $.

For a square system at $T= 2 \, T_c $ and all other parameters corresponding to the critical point, $c$ can be readily extracted from the quantity
\begin{equation}
\label{c}
 I_2 (\ell, L) - I_2 (L/2 , L)
 =\mathcal{J} (c ) \equiv
\frac{c} {2}
\ln \left(
\frac {  f \left ( \ell / L \right )  \, f \left (1- {  \ell} /L  \right ) }
% f \left (1- \frac{  \ell} {L}  \right ) }
{  f^2 (1 /2) }
\right ) .
\end{equation}
%%%%%%%%%%%%
We compute $ [  I_2 (\ell, L) - I_2 (L/2 , L) ]$ numerically using Monte Carlo simulations
at $T= 2 \, T_c $ and compared our data with this theoretical expectation. 
Of course, the numerical data thus obtained is affected by significant finite-size effects. Hence, in order
to use the above expression to obtain $c$, we perform a finite-size extrapolation as described in the next section.

%%%%%%%%%%%%%%%%%%%%%%%%%%%
{\em Results --}
%%%%%%%%%%%%%%%%%%%%%%%
\begin{figure} 
\centering
\includegraphics[width=0.47 \textwidth]{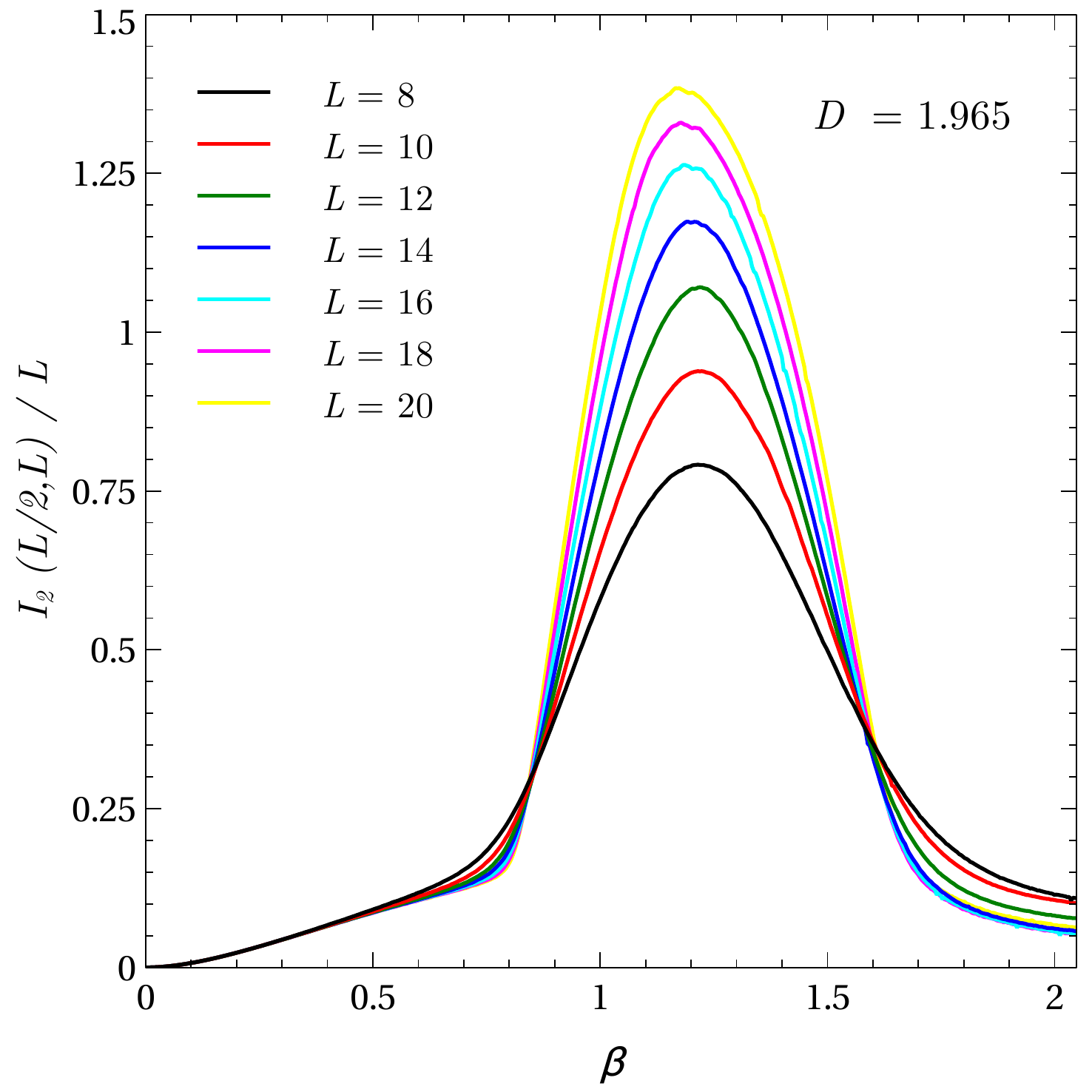}
\caption{\label{i2}
(color online). The RMI per boundary length, $I_2 (L/2 , L) / L $, as a function of $   \beta $.
Crossings are seen at $   \beta_c  /2 $ and $   \beta_c  $.}
\end{figure}
%%%%%%%%%%%%
%%%%%%%%%%%%%%%%%%%%%
We know that the Blume-Capel model has a line in the $(D,   T)$ plane representing a second-order (or continuous) phase transition, which terminates at the tricritical point, meeting another line corresponding to a first-order phase transition \footnote{These two lines, in spite of representing phase transitions of different orders, actually form one continuous line separating the ordered phase from the disordered (or paramagnetic) phase.}. 
This line of phase transitions can be detected by the second R\'enyi entropy.
As an example,
Fig.~\ref{i2} shows the RMI as a function of temperature for $D  = 1.965$,
revealing a transition at $T_c$ and $2 \,T_c$ as crossings in $I_2(L/2, L)  /L$.
The data used for this plot has been obtained by thermodynamic integration and imposing periodic boundary conditions on the lattice.
Although the RMI curve looks continuous, 
the fact that these parameters are inconsistent with the tricritical point can be checked by calculating the central charge from GMI using a finite-size extrapolation.

Let us elaborate on how this finite-size analysis can be implemented.
The function $ y(1/L) = m  /L \,  +  \, c_{ \text{extr} } $ is obtained for each $ \ell/L$, where $m$ is the slope
and $ c_{ \text{extr}  }$ is the $y$-intercept for the data points corresponding to $ \Big ( x\equiv 1/L , \, y(x) \equiv  I_2 (\ell, L) - I_2 (L/2 , L) \Big  )$. 
After collecting the $\lbrace c_{ \text{extr} } \rbrace $ for all ratios $\lbrace \ell/L \rbrace$, we fit
these with the fitting function $\mathcal{J}(c )$, keeping $c$ as the free parameter.
%%%%%%
In other words, the set $\lbrace \ell/L, c_{ \text{extr} } \rbrace$ is fitted by numerically searching for the value of $c$ which makes $\mathcal{J}(c )$ fit the data best.
In addition, we have computed the $\chi^2$ estimates, which
tell us how close the numerically extracted values are to the theoretical prediction of $c = 0.7$.

%%%%%%%%%%%%%%%%%%
 \begin{figure}
 \centering
 \subfloat[][]{\includegraphics[width=0.48 \textwidth]{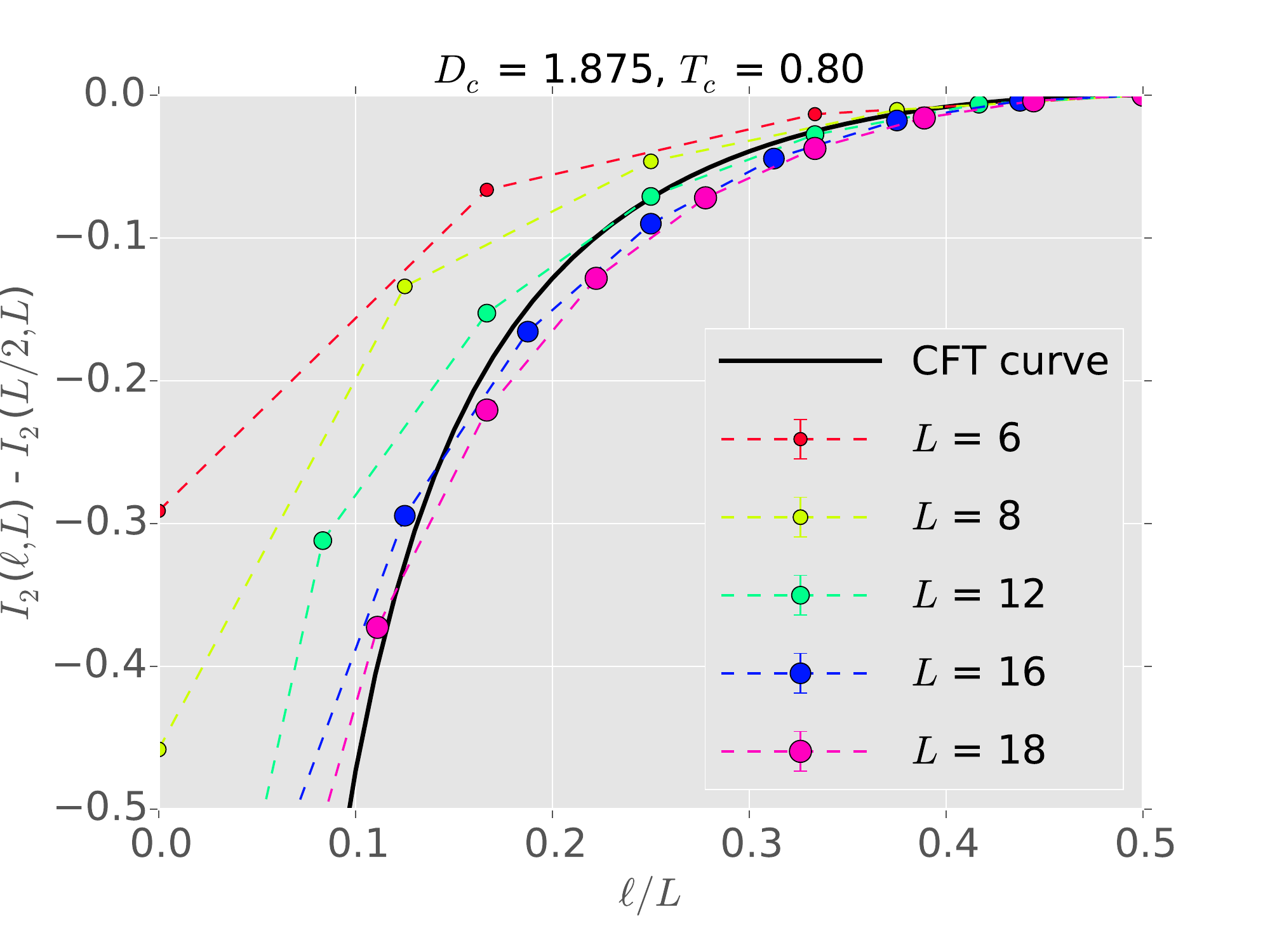}}\\
 \subfloat[][]{\includegraphics[width=0.48\textwidth]{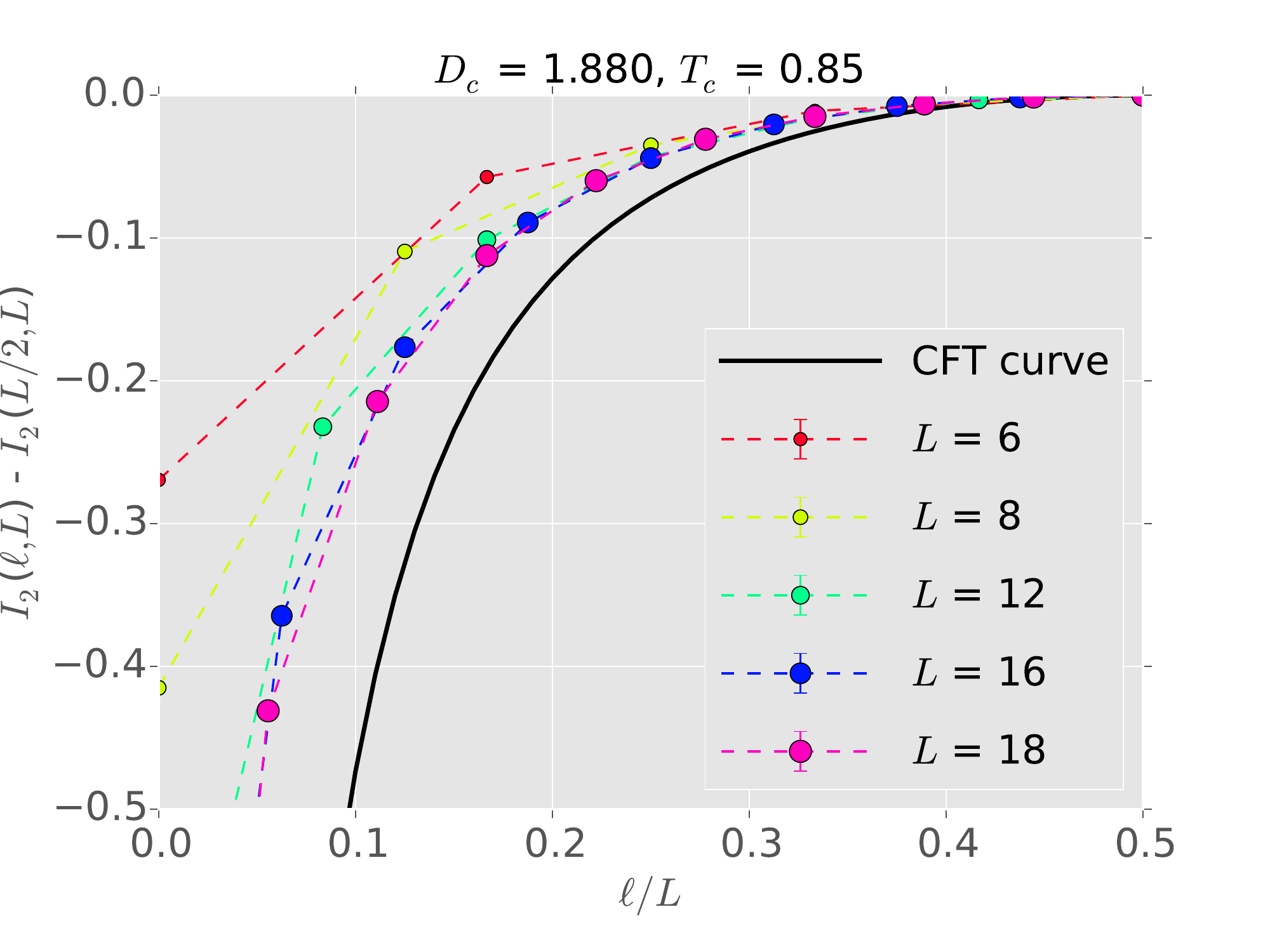}}
 \caption{\label{i2-not-c}
 (color online). Representative curves for values of $ ( D_c , T_c  )$ which are off-critical and hence do not give the central charge close to the theoretical value of $0.7 $ at the tricritical point. The ``CFT curve" corresponds to the plot of $\mathcal{J}(0.7)$.}
 \end{figure}
%%%%%%%%%%%%%%%%%%%%%%

\begin{figure}
\centering
\subfloat[][]{\includegraphics[width=0.48\textwidth]{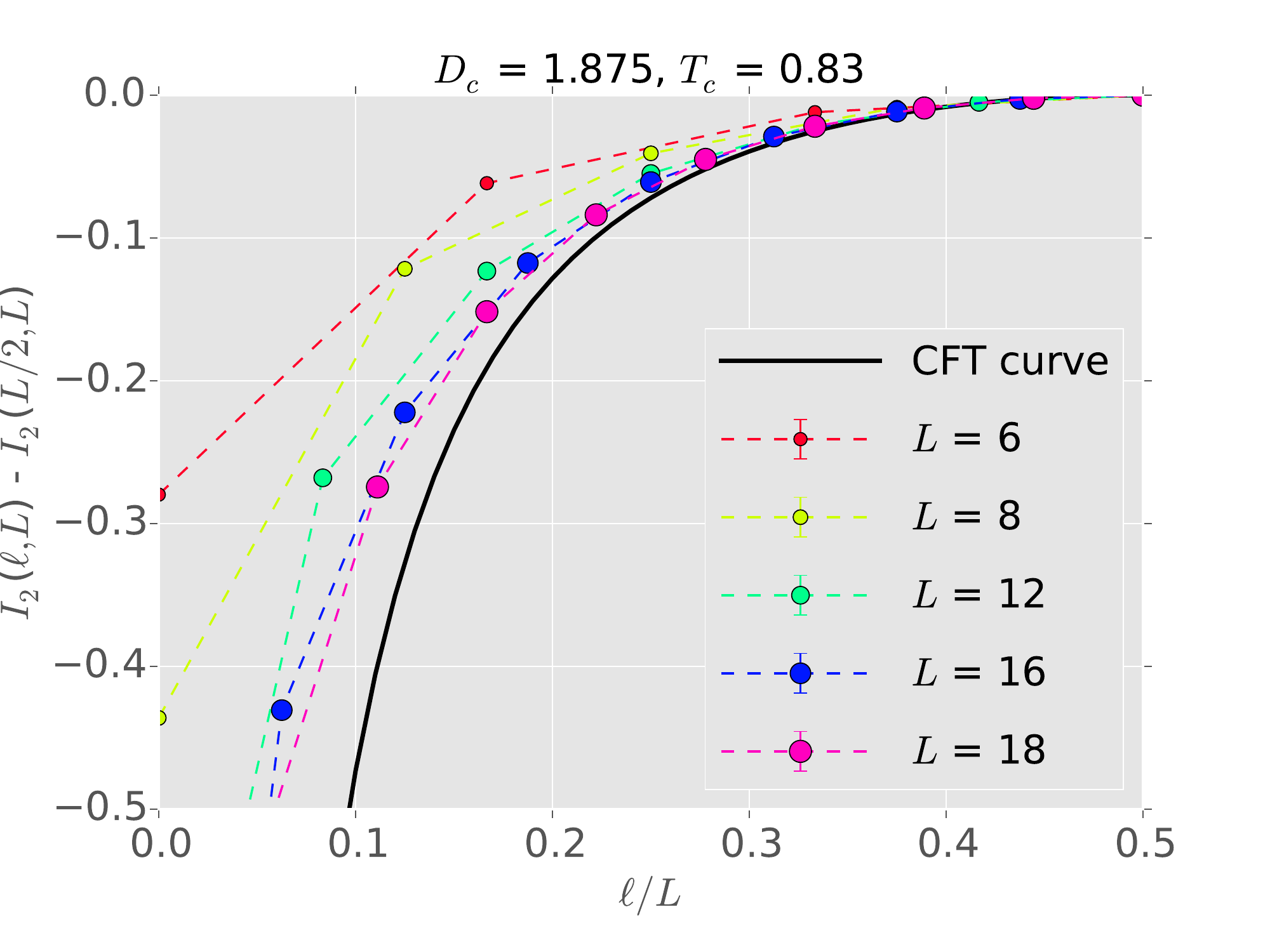}}\\
\subfloat[][]{\includegraphics[width=0.48\textwidth]{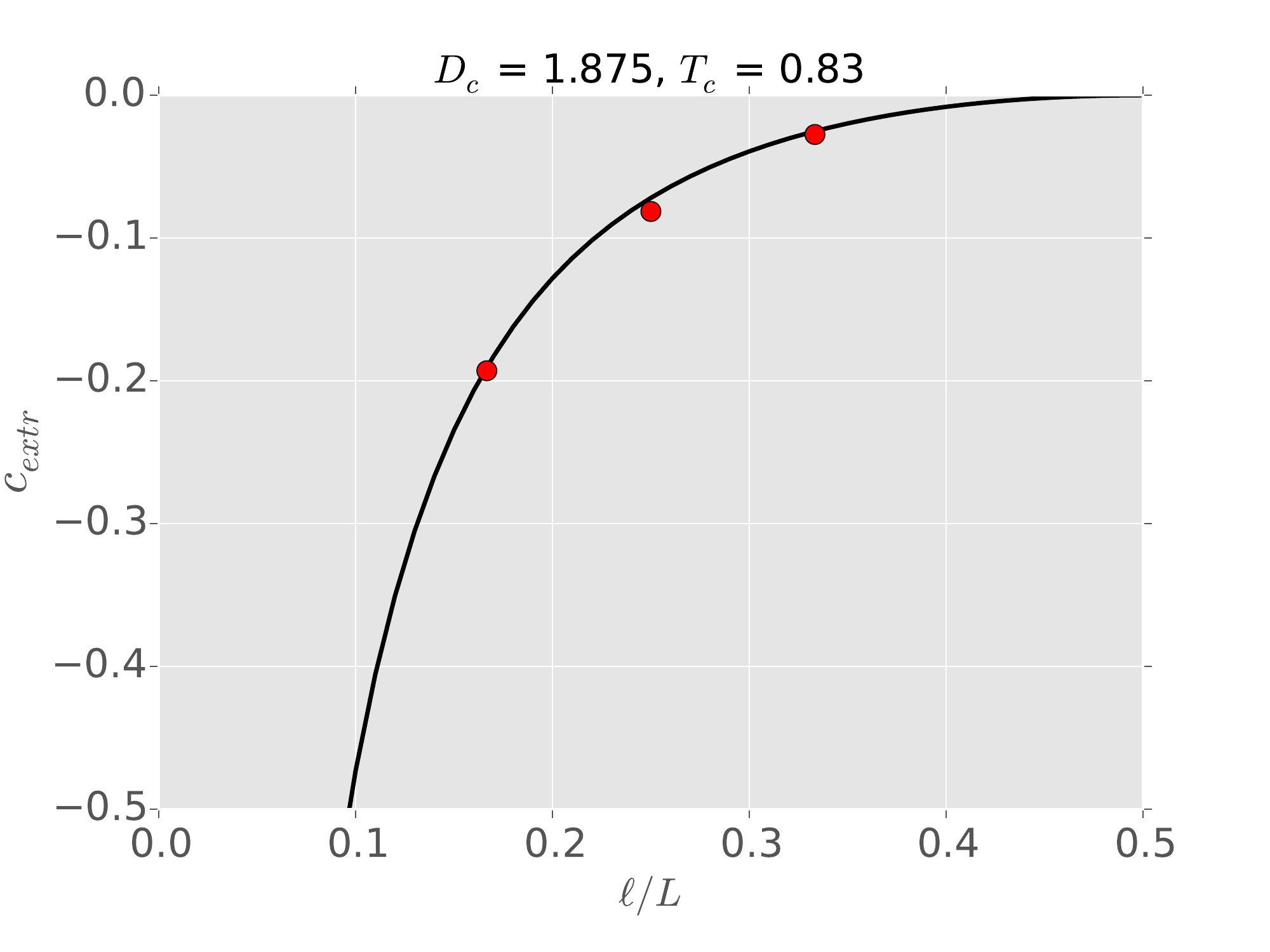}}
\caption{\label{i2-for-c}
(color online). The first plot shows the data points obtained for $ ( D_c  = 1.875 ,  T_c = 0.83 )$, lying within the region of least $\chi^2$, with the ``CFT curve" corresponding to $\mathcal{J}(0.7)$. The second plot shows that the points, obtained after finite-size extrapolation, lie on the CFT curve.
}
\end{figure}

For the Blume-Capel model, the GMI at the critical point depends on two parameters: $D_c $ and $ T_c  $.
This makes it harder to pin down the tricritical point compared to the models studied in earlier works by this same technique \cite{stephen2013,stephan2014}, for which the critical point  corresponding to a CFT were dependent only on the parameter $ T_c  $. We have scanned the parameter space to find the tricritical point.
Fig.~\ref{i2-not-c} shows the behavior of $c$  as a function of the ratio $\ell /L$, for two sets of $ ( D_c ,  T_c  )$, which do not give a central charge consistent with the theoretical tricritical value of $0.7$, after fitting $
\lbrace c_{ \text{extr} } \rbrace $ with $\mathcal{J}(c)$. This illustrates how we may discard values of $(D,  T)$ which have no possibility of corresponding to the tricritical point. 

To obtain an estimate for the tricritical parameters $  T_{tc}$ and $D_{tc}$,
we have implemented a two-step procedure as follows. In the first step, the best-fit $c$ has been calculated from $\lbrace 
c_{\text{extr} }\rbrace $, as described above, for a range of sizes $\lbrace 6,8,12,16,18 \rbrace $ and aspect ratios $
\lbrace 1/4, 1/6, 2/6, 1/8, 3/8 \rbrace $.
In the second step, we assume $c$ is fixed to $0.7$, and calculate a goodness-of-fit measure $\chi^2$. 
This $\chi^2$-estimate, which quantifies the quality of our fits to $c=0.7$, is crucial in determining a region of critical parameters which gives our best-fit to the tricritical CFT form.
%%%%%

\begin{figure}
 \captionsetup[subfigure]{labelformat=empty}
 \centering
 \subfloat[][]{\includegraphics[width=0.45\textwidth]{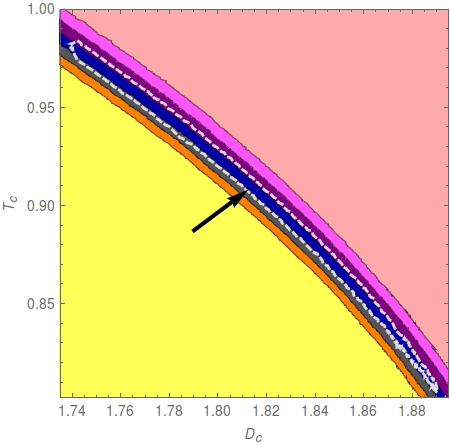}}
 \vspace{0.0001mm}
 \subfloat[][]{\includegraphics[width=0.4 \textwidth]{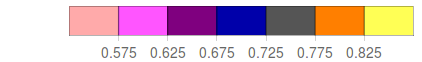}}
 \caption{\label{region}
 (color online). The contourplot shows the $c$-contours in the $( D_c   ,T_c )$-plane, obtained by fitting the data to the universal shape function $\mathcal{J} (c)$ after finite-size extrapolation. The region within
 $  D_c    \in  \lbrace 1.740 ,  1.895 \rbrace   $ and $ T_c  \in   \lbrace 0.80 ,  0.97 \rbrace $ gives the closest match to the actual $c=0.7$. The dashed silver line, highlighted by the arrow, encloses the region with least $\chi^2 $ (in arbitrary units), assuming fitting to $ c= 0.7 $.}
 \end{figure}

To obtain our estimate of the tricritical point, we have restricted our search to the region
$  D_c   \in  \lbrace 1.735 ,  1.895 \rbrace   $ and $ T_c  \in   \lbrace 0.80 ,  1.00 \rbrace $, taking help from previous studies in the literature \cite{beale,yusuf}, in order to make a judicious choice of the parameter ranges.
Varying $\left ( D_c   ,  T_c   \right )$, and repeating our finite-size extrapolation and $\chi^2 $ computation, we have found numerically that values consistent with 
$c=0.7$ are obtained roughly within this parameter range.
Fig.~\ref{i2-for-c}(a) shows $ [ I_2 (\ell, L) - I_2 (L/2 , L)   ] $ obtained for a representative point $ ( D_c = 1.875 ,T_c = 0.83 )$, lying within the region of minimum $\chi^2$. Fig.~\ref{i2-for-c}(b) shows how well the $\lbrace  c_{ \text{extr} } \rbrace $, obtained from the finite-size extrapolation of this data set, fall on the expected CFT curve.
This analysis can be summarized in the contour plot in Fig.~\ref{region}, which illustrates the fitted values of $c$, together with
an outline of the valley that occurs in the $\chi^2$ measure for $c=0.7$.  The overlap gives us a range for our estimated value of $ T_{tc}$ and $ D_{tc}$.

%%%%%%%%%%%%%%%%%%

%%%%%%%%%%%%%%%%%%%%%%%%%%%%%%%%%%%%%%%%%%%%%%%%%%%%%
{\em Discussion --}
In this paper, we have estimated the position of the tricritical point for the spin-1 classical Blume-Capel model on a square lattice by using classical Monte Carlo calculations of second R{\'e}nyi entropy.  
First, by analyzing the Geometrical Mutual Information (GMI), 
we have calculated the central charge of the low-energy conformal field theory (CFT) 
description of the critical point, and confirmed that it agrees with the known theoretical value of $c=0.7$.
Then, restricting our range of model parameters by looking at the best-fit of the data to 
this value, we have obtained a range of coupling constants consistent with the tricritical point.
Our technique is an interesting demonstration of the power of the GMI to distinguish tricritical CFTs numerically, without
reliance on an order parameter or thermodynamic observable.
However, determination of the R\'enyi entropy requires a calculation of a ratio of partition functions, which can only be obtained
through thermodynamic integration, or variations of a ``ratio-trick''.  Thus, the system sizes obtained will be smaller than what is 
possible with conventional estimators.  
Attempting to control these finite-size effects with careful extrapolations leads to the conclusion that
the tricritical point can be anywhere within the minimum $\chi^2 $ region of Fig.~\ref{region}, lying in the range $  D_c    \in  \lbrace 1.735 ,  1.895 \rbrace   $ and $  T_c   \in   \lbrace 0.80 ,  0.97 \rbrace $.   This region is consistent with the previous surveys in literature, obtained through a 
variety of other numerical techniques \cite{yusuf}. This shows that despite difficulties related to limitations imposed by finite-size effects, it is relatively straightforward to obtain a reasonable estimate of the tricritical point from our R\'enyi entropy data. 

For the model studied in this paper, where the central charge is known, 
we have demonstrated how its knowledge can be used to provide an estimate of two parameters
$(D_c, T_c)$ which define the tricritical point.
The same procedure could be applied in other interesting systems, where the order parameter is not 
known, but which contain critical lines with well-defined $c$-values \cite{affleck2016}.
In other higher-dimensional models, where the system cannot be mapped to $(1+1)$-dimensional CFTs, 
it would be interesting to extend the definition of the GMI to explore higher-dimensional analogues of $c$, if  the values of critical 
parameters are known.
This would further establish the diverse utility of information theory techniques in the arena of statistical mechanics and condensed-matter theory.

%%%%%%%%%%%%%%%%%%%%%%%%%%%%%%%%%%%%%%%%%%%%%%%%%%%%%%%%%%
{\em Acknowledgments --} We thank P.~Fendly for enlightening discussions. I.M. is grateful to A. Bhattacharya and
 L.~E.~Hayward Sierens for helping with the basics of coding.
 This work was made possible by the computing facilities of SHARCNET. Support was provided 
by NSERC of Canada (I.M. and R.G.M.), the Templeton
Foundation (I.M.), the FP7/ERC Starting Grant No. 306897 (S.I.), 
and the National Science Foundation under Grant No. NSF PHY11-25915 (R.G.M).
Research at the Perimeter Institute is supported, in
part, by the Government of Canada through Industry Canada
and by the Province of Ontario through the Ministry of
Research and Information.

\bibliography{bc-draft}

\end{document}